\documentclass[12pt]{iopart}
\usepackage{epsfig}

\def\laq{\raise 0.4ex\hbox{$<$}\kern -0.8em\lower 0.62ex\hbox{$\sim$}}
\def\gaq{\raise 0.4ex\hbox{$>$}\kern -0.7em\lower 0.62ex\hbox{$\sim$}}

\newcommand{\beq}{\begin{equation}} 
\newcommand{\eeq}{\end{equation}}
\newcommand{\bea}{\begin{eqnarray}} 
\newcommand{\eea}{\end{eqnarray}}

\begin{document}

\title[Gravitational waves from inspiraling 
binary black holes]{Gravitational waves from inspiraling binary black holes }

\author{Alessandra Buonanno\dag \ddag}

\address{\dag Institut d'Astrophysique de Paris (GReCO, FRE 2435 du CNRS), 
98$^{\rm bis}$ Boulevard Arago, 75014 Paris, France} 
\address{\ddag Theoretical Astrophysics and 
Relativity Group, California Institute of Technology, 
Pasadena, 91125, CA, USA}

\begin{abstract}
Binary black holes are the most promising candidate sources for 
the first generation of earth-based interferometric 
gravitational-wave detectors. 
We summarize and discuss the state-of-the-art analytic 
techniques developed during the last years to better describe 
the late dynamical 
evolution of binary black holes of comparable masses. 
\end{abstract}

\ead{buonanno@tapir.caltech.edu}

\vspace{8cm}

\noindent
{\small Summary talk given at the $4^{th}$ Edoardo Amaldi 
Conference on Gravitational Waves, Perth, Australia, 8-13 July 
2001, to appear in Special Issue Article of Classical and Quantum Gravity}
\maketitle

\section{Introduction}
\label{sec1}
Binary systems made of compact objects (neutron stars or black holes) 
that spiral in toward coalescence because of gravitational-radiation damping 
are among the most promising candidate sources for 
interferometric gravitational-wave (GW) detectors, such as 
the Laser Inteferometric Gravitational Wave Observatory (LIGO),  
VIRGO, GEO and TAMA~\cite{Inter}. The inspiral waveform will enter the detector 
frequency band during the last few minutes or seconds of evolution of the binary 
and the GW community plans to track the signal phase and build up 
the signal-to-noise ratio by integrating the signal for the 
time during which it stays in the detector bandwidth. 
This is achieved by filtering the detector output with a 
template which is an (approximate) theoretical copy of the 
observed signal. 

Einstein theory predicts that the radiative transverse traceless (TT) 
gravitational field $h_{i j}^{\rm TT}$, far away from the source, 
is related to the motion and the structure of the source, at lowest 
order in the post-Newtonian (PN) expansion, by the quadrupole formula
\beq
h_{i j}^{TT}(T, D) = \frac{2 G}{c^4 D}\,{\cal P}_{ijkm}(\mbox{\boldmath{$N$}})
\,\frac{d^2}{dT^2}{Q}_{k m} \left (T - 
\frac{D}{c} \right )\,, 
\label{1}
\eeq
where $Q_{i j}$ ($i,j = 1,2,3$) is the tracefree quadrupole moment of 
the source; $D$ is the distance from the source; $\mbox{\boldmath{$N$}}=
\mbox{\boldmath{$X$}}/D$ is the unit vector from the source 
to the observer; ${\cal P}_{ijkm}(\mbox{\boldmath{$N$}})$ is the TT projection 
operator onto the plane orthogonal to $\mbox{\boldmath{$N$}}$;
$G$ is the Newton constant; and $c$ is the speed of light.
{}From Eq.~(\ref{1}) (and its extensions at higher PN orders)  
we see that the more precisely we know the two-body 
motion, the more accurately the PN template $h^{\rm TT}_{i j}$ will  
describe the ``real'' gravitational waveform.

\begin{figure}
\begin{center}
\vspace{-0.3cm}
\includegraphics[width=5.5cm,angle=-90]{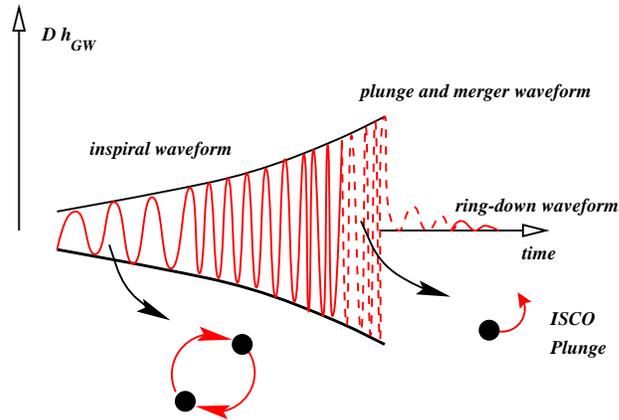}
\caption{\sl Typical gravitational waveform emitted 
throughout the inspiral, plunge and ring-down phases.} 
\label{Fig3}
\vspace{-0.7cm}
\end{center}
\end{figure}

In Fig.~\ref{Fig3} we show a typical gravitational waveform.
The part of the waveform drawn with a continuous line 
is emitted during the inspiral phase when the two 
black holes are largely separated, e.g., $ r \geq 10 GM/c^2$  
where we denoted by $r$ the radial separation and by $M$ the total mass 
of the binary system. During the inspiral, the two black holes 
follow an adiabatic sequence of quasi-circular orbits. 
The equation of motion in the center-of-mass frame can be written 
schematically as 
\beq
\hspace{-1.5cm}\frac{d^2 \mbox{\boldmath{${x}$}}}{d t^2} = 
- \frac{G M \mbox{\boldmath{${x}$}}}{r^3}\,
[ 1 + {\cal O}({{\epsilon}}) + 
{\cal O}({{\epsilon^2}}) + 
{\cal O}({{\epsilon^{5/2}}}) + {\cal O}({{\epsilon^{3}}})
+ \cdots ] \times [ 1 + {\cal O}({{\nu}}) + {\cal O}({{\nu^2}}) 
+ \cdots ]\,,\,
\label{2}
\eeq
where $\mbox{\boldmath{${x}$}}$ denotes the separation vector  
between the two bodies and $r= |\mbox{\boldmath{${x}$}}|$.
Equation (\ref{2}) is characterized by a double expansion: in 
the PN parameter $\epsilon \sim 
{v^2}/{c^2} \sim GM/(c^2 r)$, and in the parameter $\nu = m_1\,m_2/M^2$, 
where $m_1$ and $m_2$ are the masses of the two black holes.
The parameter $\nu$ ranges between $0$ (test-mass limit) 
and $1/4$ (equal-mass case).

It is well known that the PN expansion converges badly: 
as the two bodies draw closer, and enter the nonlinear, strong curvature 
phase, the motion becomes relativistic, e.g., $v/c \sim 0.3$, 
and it is more and more difficult to extract reliable 
information from the PN series. More specifically, when the distance 
between the inspiraling black holes shrinks to $ r \,\laq\, 10 GM/c^2$, 
the PN expansion can no longer be trusted~\cite{BCT98}.
The dashed line in Fig.~\ref{Fig3} depicts the 
(less known) part of the waveform emitted during the final phase 
of evolution, when nonlinearities and strong curvature 
effects become important and nonperturbative analytical and/or 
numerical techniques should be used to describe it. 
This final phase includes the transition 
from adiabatic inspiral to plunge, beyond which the two-body 
motion is driven (almost) only by the conservative 
part of the dynamics. For nonspinning binary black holes, 
the plunge starts at the innermost stable circular orbit (ISCO) 
of the binary black holes. Beyond the plunge the two black holes  
merge, forming a Kerr black hole. As the system reaches the 
stationary Kerr state, the nonlinear dynamics of the merger 
resemble more and more the oscillations of the black-hole quasi-normal 
modes~\cite{QNR} and the gravitational signal will be a superposition 
of exponentially damped sinusoids (ring-down waveform). 
However, if the black holes carry a big  
spin, it is likely that the plunge stage will never 
be reached and the late dynamical phase will be much 
more complicated in this case. 

Likely, the first detection of gravitational waves with LIGO/VIRGO 
interferometers will come from binary systems made of black holes 
of comparable masses, say with a total mass $M \simeq 10 M_{\odot} 
\div 30 M_{\odot}$. In fact, if we assume binary black holes 
are originated by ``population synthesis''~\cite{postnov}, 
the estimated detection rate per year is
$\laq \, 4 \times 10^{-3} -0.6$ at $100\, {\rm Mpc}$~\cite{KT}, 
while if globular clusters are considered as 
``machines'' for making binary black holes~\cite{PZ99}
the detection rate per year is $\sim 0.04 -0.6$ at $100\, {\rm Mpc}$~\cite{KT}.
These numbers, especially the latter, are more optimistic 
than the estimated detection rate per year for neutron 
star binaries, which is $\laq\, 3 \times 10^{-4} -0.3$ at $20 \,{\rm Mpc}$~\cite{KT} 
or for neutron star/black hole binaries, which 
is $\laq\, 4 \times 10^{-4} -0.6$ at $43 \,{\rm Mpc}$~\cite{KT}.

Although the number of cycles in the LIGO/VIRGO frequency band of    
gravitational waves emitted by comparable-mass binary black holes 
is not very high, on the order of $10 - 200$, 
however these particular sources demand a more careful analysis 
because the gravitational waves the detectors will be 
sensitive to are emitted during the final stages of inspiral  
where PN expansion fails. For example, in the nonspinning 
case the GW frequency at the ISCO (evaluated using the 
Schwarzschild ISCO) is $f_{\rm GW}^{\rm ISCO} 
\simeq 220\, {\rm Hz}$ for $M=20 M_\odot$ and 
$f_{\rm GW}^{\rm ISCO} \simeq 167\, {\rm Hz}$ for $M=30 M_\odot$, 
well inside the LIGO/VIRGO  band.  
Moreover, black holes could carry big spins which could affect the waveforms, so 
if data analysis will be done with  spinless templates 
there is considerable chance to miss the gravitational-wave signal.  

In the next section we shall summarize what has been done in 
the literature to cope with those problems and which issues 
remained to be solved.

\section{Analytic methods to predict gravitational waveforms}
\label{sec2}

For spinless black hole binaries, we consider the so-called restricted 
waveform: $h(t)=v^2\,\cos (\varphi_{\rm GW}(t))$, 
where $\varphi_{\rm GW} = 2 \varphi$ with $\varphi$ the orbital phase 
and $v$ is the invariantly defined velocity $v=(M\,\dot{\varphi})^{1/3} 
= (\pi M\,f_{\rm GW})^{1/3}$.  
This waveform is obtained disregarding all the multipolar components
appearing in the gravitational waveform except the quadrupolar one 
[see Eq.~(\ref{1})]. 
To determine in the adiabatic limit the evolution of the GW phase, 
$\varphi_{\rm GW}(t)$, it is sufficient to use the energy-balance 
equation 
\beq
\frac{d {\cal E}}{d t} = - {\cal F}\,,
\label{3}
\eeq
relating the orbital energy function ${\cal E}$ (center-of-mass 
energy that is conserved in absence of radiation reaction) to 
the gravitational-flux (or luminosity) function ${\cal F}$, which  
are known for quasi-circular orbits as a PN expansion in $v$.
It is easily shown that Eq.~(\ref{3}) is equivalent to the 
following system of differential equations [see, e.g.,~\cite{DIS1}] 
\beq
\frac{d \varphi_{\rm GW}}{d t} = \frac{2v^3}{M}\, \quad \quad  
\frac{d v}{dt} = - \frac{{\cal F}(v)}{M\,d{\cal E}(v)/dv}\,,
\label{4}
\eeq
whose solution provides the phasing $\varphi_{\rm GW}(t)$ during the inspiral.

\subsection{``Genuine'' post-Newtonian calculations}
\label{subsec2.1}
Let us summarize what we know about the two crucial ingredients 
${\cal E}$ and ${\cal F}$ entering Eq.~(\ref{4}). 
The equations of motion of two compact bodies at 2.5 PN 
approximation were first derived in Refs.~\cite{DD}. The 3PN equations 
of motion have been obtained by two separate groups: 
Damour, Jaranowski and Sch\"afer~\cite{DJS} used the Arnowitt-Deser-Misner 
(ADM) canonical approach while Blanchet, Faye, and de Andrade~\cite{DBF} worked with 
the PN iteration of Einstein equations in harmonic gauge. Recently, 
Damour et al.~\cite{DJSd} working in the ADM formalism and  
applying dimensional regularization, uniquely determined the so-called 
``static'' parameter which entered the 3PN equations of 
motion~\cite{DJS,DBF} and up till then was unknown. 
Thus, at present time the energy function ${\cal E}$ is known up to 3PN order.

The gravitational flux emitted by compact binaries was 
first computed at 1PN order in Ref.~\cite{1PN}. Subsequently, 
it was determined at 2PN order using a formalism based 
on multipolar and post-Minkowskian approximations 
and independently by using a direct integration of the 
relaxed Einstein equations~\cite{2PN}. Non-linear effects of 
tails at 2.5PN and 3.5PN orders were computed in Refs.~\cite{2.5PNand3.5PN}. 
More recently, the gravitational-flux function for quasi-circular 
orbits has been derived up to 3.5PN order~\cite{BIJ}. However, at 3PN order~\cite{BIJ} 
the gravitational-flux function depends on an arbitrary parameter which is 
not fixed by the regularization scheme used by the authors.

Although the knowledge of higher order PN corrections to 
${\cal E}$ and ${\cal F}$ is a necessary ingredient 
to extract the phase of gravitational signals emitted 
by neutron star or black-hole/neutron-star binaries, however  
it is not by itself sufficient for computing GWs of comparable-mass 
binaries. As underlined above, this is due to the fact that for these 
sources LIGO/VIRGO will detect gravitational signals emitted 
when the motion is relativistic and (genuine) perturbative PN calculations can 
no longer be trusted.

\subsection{Post-Newtonian resummation methods} 
\label{subsec2.2}

Certainly, the best way of extracting the GW signal emitted 
by comparable-mass binaries during the last stages of inspiral,  
would be to solve numerically the Einstein equations of a binary black hole system.
Unfortunately, despite the interesting progress made by the numerical relativity 
community during the recent years~\cite{C,TB,PTC,GGB,NR}, 
an estimate of the waveform emitted by black-hole binary has not 
yet been provided. Preliminary results for the plunge, 
merger and ring-down waveforms were only recently 
obtained~\cite{BBCLT01} and they use initial conditions at the ISCO 
which differ from PN predictions. To overcome this gap and 
tackle the delicate issue of the late 
dynamical evolution, various \emph{nonperturbative analytical} approaches 
have been proposed~\cite{DIS1,BD1,BD2,EOB3PN} to study the motion of 
two spinning and nonspinning bodies in general relativity.  

The main features of the various PN resummation methods can be 
summarized as follows: (i) they provide an analytic 
(gauge-invariant) resummation of the 
orbital energy function ${\cal E}$ and gravitational flux function 
${\cal F}$, (ii) they can describe the motion (and provide 
the gravitational waveform) beyond the adiabatic approximation 
and (iii) they can in principle be extendible to higher 
PN orders. More importantly, they can be used to provide initial data for black 
holes just starting the plunge motion which can be 
used by the numerical relativity community to evolve  
the full Einstein equations during the merger phase. 
However, the resummation methods are also based on some assumptions 
hard to prove rigorously -- for example, in deriving 
the orbital energy and the gravitational flux functions 
in the comparable-mass case, it is assumed that they are smooth deformations 
of the analogous quantities in the test-mass limit. 
Moreover, in absence of an exact solution or of experimental data we can test 
the robustness and reliability of those resummation approaches only using 
internal convergence tests. 

As underlined at the beginning of Sec.~\ref{sec2}, in absence 
of spins, the two crucial ingredients necessary to extract the 
GW phasing are the orbital energy 
function ${\cal E}$ and the gravitational flux function ${\cal F}$. 
In Sec.~\ref{subsubsec2.2.1} we shall discuss 
a resummation approach which provides a better behaved flux-type function~
\cite{DIS1}, while in Sec.~\ref{subsubsec2.2.2} we shall summarize 
another resummation method which provides an improved 
energy-type function~\cite{BD1,EOB3PN}. 
By combining these two resummation methods, Buonanno and Damour~\cite{BD2} 
proposed a way of describing the black hole motion beyond the adiabatic limit, 
including the transition from inspiral to plunge, whose main features 
are discussed in Sec.~\ref{subsec2.3}.

\subsubsection{Pad\'e approximants}
\label{subsubsec2.2.1}

Starting from the PN expansions of ${\cal E}(v)$ and ${\cal F}(v)$ Damour, Iyer 
and Sathyaprakash~\cite{DIS1} proposed a new class of waveforms based 
on the systematic use of Pad\'e resummation,
\footnote{~Let us assume that we know the function $g(v)$ only through 
its Taylor approximant $G_N(v) = g_0 + g_1\,v + \cdots + g_N\,v^N \equiv T_N[g(v)]$. 
The idea of Pad\'e summation \cite{BO} is to replace the power series $G_N(v)$ by the sequence 
of rational functions $$P_K^M[g(v)] = \frac{A_M(v)}{B_K(v)} \equiv
\frac{\sum_{j=0}^M a_j\,v^j}{\sum_{j=0}^K b_j\,v^j}$$
with $M+K=N$ and $T_{M+K}[P_K^M[g(v)]]=G_N(v)$ ($b_0=1$).
For $M, K \rightarrow + \infty$, $P_K^M(v) \rightarrow g(v)$.
The choice $K=M$ or $K=M+1$ defined the near diagonal Pad\'e approximants.}
which is a standard mathematical 
technique used to accelerate the convergence of poorly converging or even 
divergent power series. 
For lack of space we shall discuss briefly only the Pad\'e approximant 
to the flux function ${\cal F}$. 

In the test mass limit ($\nu \rightarrow 0$) the flux function has a simple 
pole at the light ring ($v^2=1/3$). Damour et al.~\cite{DIS1} argued 
that the origin of this pole is quite general and that we should expect 
a pole singularity also in the equal-mass case. Thus, after factoring out 
this pole (defined by what they call the better behaved energy-function $e(v)$, 
see Ref.~\cite{DIS1}), they introduced the better behaved flux-function 
$f(v) \equiv (1-v/v_{\rm pole})\,{\cal F}(v)$. Using the 2.5PN expansion 
of ${\cal F}(v)$ and applying the (near diagonal) Pad\'e 
resummation \footnote{~See previous footnote.} 
to $f(v)$, they derived the $v^n$-Pad\'e approximant to 
the gravitational flux function:
\beq
{\cal F}^n_{P}(v;\nu) = \frac{1}{1-v/v^n_{\rm pole}(\nu)}\,
{f}^n_{P}(v;\nu)\,.
\label{5}
\eeq
To test the reliability of the result, Damour et al. showed that in 
the test mass limit the (near diagonal) Pad\'e approximant to ${\cal F}(v;\nu=0)$, 
given by Eq.~(\ref{5}), 
exhibits a very good convergence toward the exact result, which is numerically 
known when $\nu=0$. [See Fig. 3 in Ref.~\cite{DIS1}.] Thus, arguing  
that the equal-mass case can be obtained as a smooth deformation 
of the test mass limit, with deformation parameter $\nu$, 
they propose ${\cal F}^n_{P}(v;\nu)$ as the 
best estimation of the GW flux for comparable-mass binaries.

\subsubsection{Effective-one-body reduction}
\label{subsubsec2.2.2}

\begin{figure}
\begin{center}
\includegraphics[width=6.5cm,angle=-90]{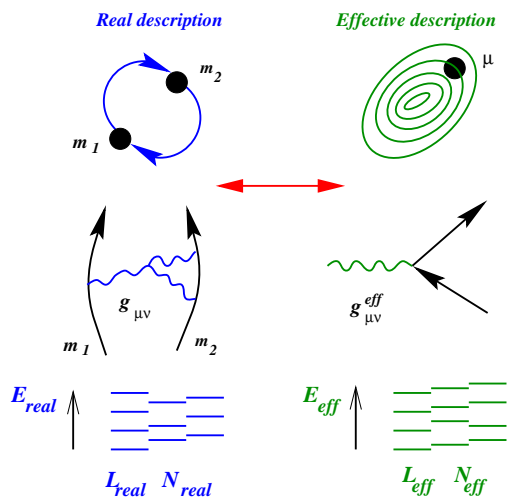}
\caption{\sl How the EOB approach matches the real two-body problem (on the left) 
and the effective one-body problem (on the right). [We defined $N = I + L$, where 
$I$ is the radial action variable and $L$ the angular momentum.]}
\label{Fig4}
\vspace{-0.7cm}
\end{center}
\end{figure}

The resummation technique discussed in this section, 
the so-called effective-one-body (EOB) approach~\cite{BD1}, 
was originally inspired by a similar approach 
introduced by Br\'ezin, Itzykson and Zinn-Justin~\cite{BIZ70} to 
study electromagnetically interacting two bodies. 
The basic idea, illustrated in Fig.~\ref{Fig4}, 
is to map the {\em real} conservative two-body dynamics up to 2PN order (see below 
for the extension at 3PN order) onto an {\em effective} one-body problem, 
where a test particle of mass $\mu=m_1 m_2/M$, with  
$m_1$, $m_2$ the black-hole masses and $M=m_1+m_2$, moves
in some effective background metric $g_{\mu \nu}^{\rm eff}$. 
This mapping has been worked out within the Hamilton-Jacobi formalism, 
by imposing that whereas the action variables of the real 
and effective description 
coincide, i.e.\ ${L_{\rm real}} = {L_{\rm eff}}$, ${{\cal I}_{\rm real}}= {{\cal I}_{\rm eff}}$, 
where $L$ denotes the total angular momentum, and 
${\cal I}$ the radial action variable, the energy axis is allowed to change, 
${{E}_{\rm real}} = f({{E}_{\rm eff}})$, where $f$ is a generic function. 
By applying the above rules defining the mapping, it was 
found that as long as radiation-reaction effects are not taken 
into account, the effective metric is just a deformation 
of the Schwarzschild metric, with deformation parameter $\nu = \mu/M$. 
The effective metric reads~\cite{BD1}:
\beq
ds_{eff}^2 = -{A(R)}\,c^2\,d t^2 + \frac{{D(R)}}{{A(R)}}\,d R^2 + R^2 \, d\Omega^2\,.
\label{6}
\eeq
More importantly the reduction to the one-body dynamics provides 
the {\it improved} real Hamiltonian
\beq
{{\cal H}_{\rm real}^{\rm improved}}(r,p_r,p_\varphi) = 
M\,c^2\,\sqrt{1 + 2\nu\,\left ( \frac{{{\cal H}_{\rm eff}^{{\nu}}}(R,P_R,P_\varphi) 
- \mu\,c^2}{\mu\,c^2}\right )}\,,
\label{7}
\eeq
where $(r,p_r,p_\varphi)$ are the ADM coordinates used in the real description. 
The effective coordinates $(R,P_R,P_\varphi)$ can be related to 
$(r,p_r,p_\varphi)$ by a (generalized) canonical transformation given 
at 2PN order in Ref.~\cite{BD1} and at 3PN order in Ref.~\cite{EOB3PN}. 
The effective Hamiltonian reads:
\beq
{{{\cal H}_{\rm eff}({{\nu}}, R, P_R, P_\varphi)}} = 
\mu\,c^2\,\sqrt{A(R)\, 
\left ( 1 + \frac{A(R)\,P_R^2}{\mu^2\,c^2\,D(r)} 
+ \frac{P_\varphi^2}{\mu^2\,c^2\,R^2} \right )}\,.
\label{8}
\eeq
We refer to the real Hamiltonian (\ref{7}) as improved real Hamiltonian
because being a (partial) PN resummation of the original 
badly convergent real Hamiltonian ${\cal H}_{\rm real} = {\cal H}_{\rm New} + 
{\cal H}_{\rm 1PN}/c^2 + {\cal H}_{\rm 2PN}/c^4 \cdots$, 
it should better capture the nonperturbative effects of the final stage 
of the black-hole motion.
Remarkably, the mapping between the real and the effective 
Hamiltonians in Eq.~(\ref{7}) coincides with the mapping
obtained in Ref.~\cite{BIZ70}
in the context of quantum electrodynamics, where these authors mapped the one-body 
relativistic Balmer formula onto the two-body energy formula. 

The EOB approach was then extended at 3PN order in Ref.~\cite{EOB3PN}. 
The authors found that starting from the 3PN level there are 
more equations to satisfy than the number of free parameters appearing 
in the energy-map and in the effective metric. Hence, they 
suggested the following two possibilities.   
At the price of modifying the coefficients of the effective metric 
at 1PN and 2PN levels, and the energy-map (\ref{7}) as well, 
it is still possible at 3PN order to (uniquely) map the real two-body dynamics onto the dynamics 
of a test mass moving on a geodesic [see for detail Appendix A of Ref.~\cite{EOB3PN}]. 
However, this solution looks quite complicated and, more importantly, 
it does not look very natural to wait to know the 3PN Hamiltonian 
to derive the matching at 1PN and 2PN levels. 
The authors then suggested to abandon the hypothesis (used at 2PN order~\cite{BD1}) 
that the effective test mass moves along a geodesic and introduced 
in the Hamilton-Jacobi equation (arbitrary) higher derivatives terms which 
provide enough coefficients to obtain the matching.   
Because of these terms the effective 3PN Hamiltonian 
is not uniquely fixed by the matching rules defined above; the general 
expression is given by Eq.~(3.12) in Ref.~\cite{EOB3PN}. 
It is interesting to note that also in the non-geodesic case 
the relation between the effective and real Hamiltonians 
is still given by Eq.~(\ref{7}).

\begin{figure}
\begin{center}
\begin{tabular}{cc}
\hspace{-0.4cm}
\epsfig{file=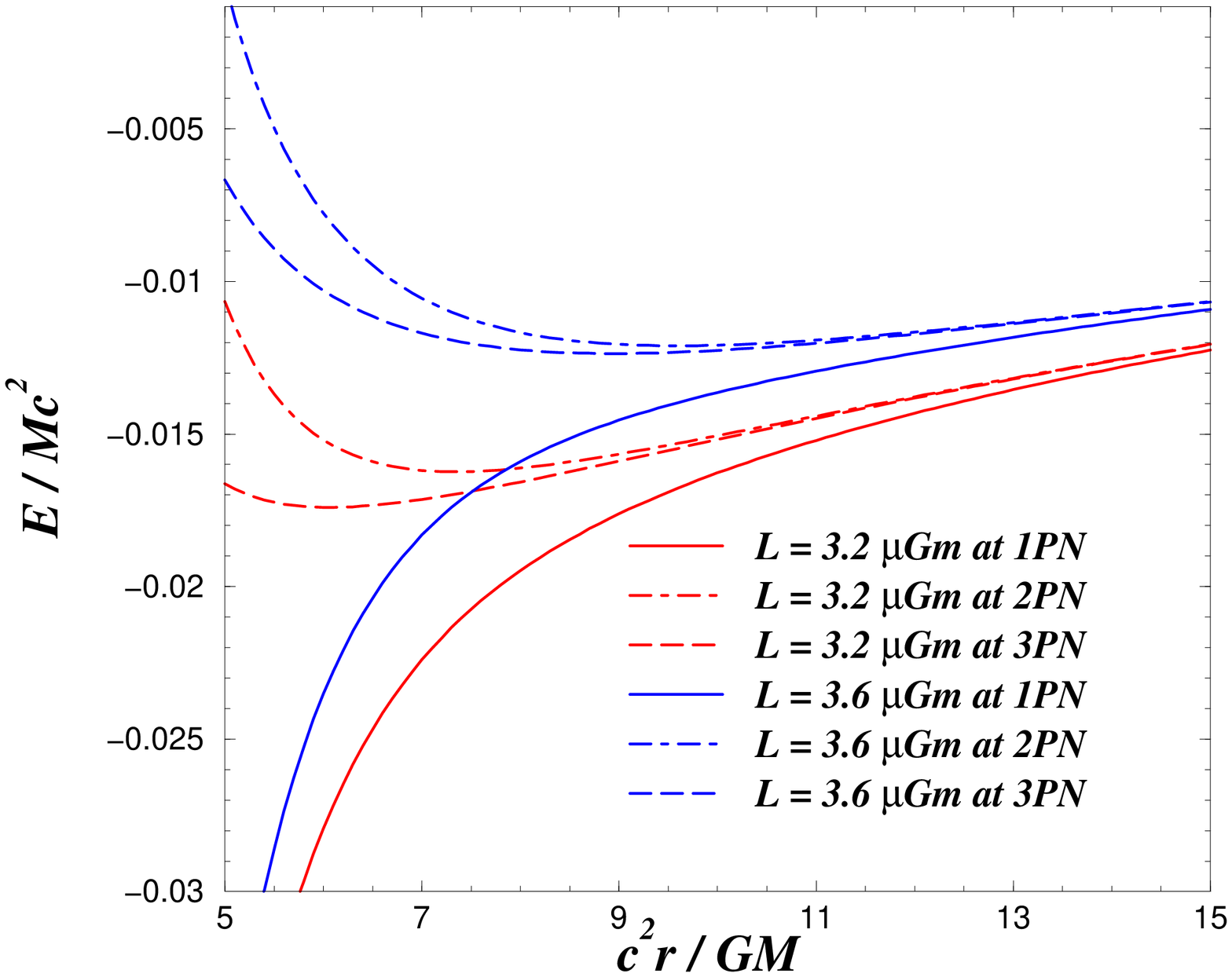,width=0.45\textwidth,height=0.5\textwidth,angle=-90} 
& \epsfig{file=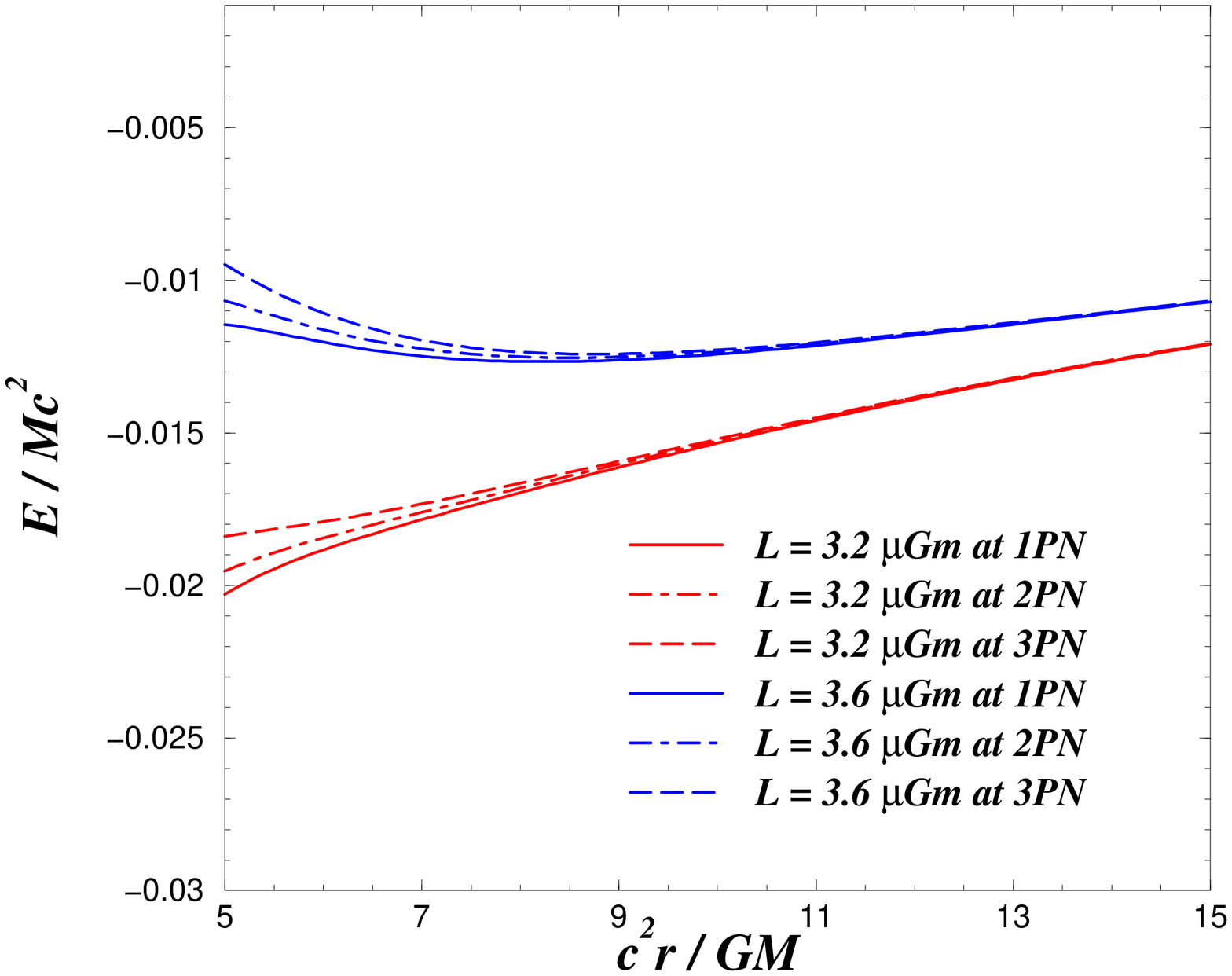,width=0.45\textwidth,height=0.5\textwidth,angle=-90} 
\end{tabular}
\caption{Comparison between the PN-expanded (on the left) and 
the EOB (on the right) binding energies for equal-mass binary.} 
\label{compa}
\vspace{-0.7cm}
\end{center}
\end{figure}

To have a qualitative understanding of the way the EOB can accelerate 
the PN convergence, we compare in Figs.~\ref{compa} the PN-expanded binding 
energy (on the left panel) to the EOB-resummed binding energy (on the right panel) 
versus the radial separation, for circular orbits, in the equal-mass case, 
for two choices of the angular momentum $L$ and at different PN orders. 
[To obtain the plots we used the 3PN-expanded
and EOB-resummed Hamiltonians given by Eqs.~(3.12) and (4.27) 
of Ref.~\cite{EOB3PN} with $\omega_s=0$~\cite{DJSd} and $z_1=0=z_2$.  
We also used the (generalized) canonical transformation between the ADM  
and the ``effective'' coordinates given by Eqs.~(4.29)--(4.36) of Ref.~\cite{EOB3PN}.]

Note that the PN-expanded energy oscillates at the various PN 
orders while the EOB energy has a monotonic behaviour. 
The fractional difference between the PN-expanded and EOB 
binding energies is, e.g., for $L = 3.2\mu GM $, 
$2.3\%$ (2PN order) and $0.24\%$ (3PN order) at $R = 9GM/c^2$ and 
$17.6\%$ (2PN order) and $2.8\%$ (3PN oder) at $R = 6GM/c^2$.

An interesting nonperturbative information of binary black hole 
systems is the presence of the ISCO defined as the solution 
of the equations $\partial {\cal H}/\partial 
r=0=\partial^2 {\cal H}/\partial r^2$. 
In the test mass limit (Schwarzschild 
metric) the ISCO exists and the nonrelativistic energy associated 
to it is ${\cal E}_{\rm isco}^{\rm Schw}= -0.01430 M\,c^2$. 
If we consider the PN-expanded real Hamiltonian [Eq.~(4.27) of Ref.~\cite{EOB3PN}] 
in the test mass limit and look for the ISCO, we find that 
at 1PN order ${\cal E}_{\rm isco}^{\rm PN-exp}= -0.00778 M\,c^2$, 
at 2PN order the ISCO does not exist, while at 3PN order 
${\cal E}_{\rm isco}^{\rm PN-exp}= -0.01129 M\,c^2$.
On the other hand, because by definition the EOB  
converges in the test mass limit to the Schwarzschild case, 
it automatically provides the correct ISCO. 
Thus, in the same way as Pad\'e approximants, 
the EOB resummation method, 
provides by construction the right prediction in the test mass limit. 
Although we cannot prove that in the comparable-mass 
case the EOB approach is converging to the right limit, 
however, we can certainly say that it shows reasonable stability 
in the predictions at 1PN, 2PN and 3PN orders.

\begin{figure}
\begin{center}
\epsfig{file=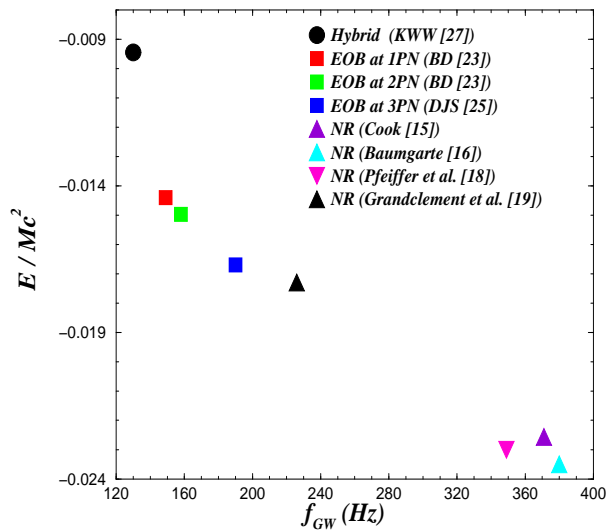,width=0.45\textwidth,height=0.5\textwidth,angle=-90}
\caption{Summary of the binding energy at the ISCO predicted 
by post-Newtonian and numerical relativity calculations for 
nonspinning binary black holes of total mass $15 M_{\odot} + 
15 M_{\odot}$.}  
\label{fig5}
\vspace{-0.7cm}
\end{center}
\end{figure}

In Fig.~\ref{fig5} we summarize the binding energy at the ISCO 
predicted by various post-Newtonian~\cite{KWW,BD1,EOB3PN} 
and numerical relativity calculations~\cite{C,TB,PTC,GGB}. 
The equal-mass ISCO binding energies predicted by the EOB 
approach at various PN orders are:
${\cal E}_{\rm isco}^{\rm EOB}= -0.01440 M\,c^2$ (1PN order),
${\cal E}_{\rm isco}^{\rm EOB}= -0.01498 M\,c^2$ (2PN order) and 
${\cal E}_{\rm isco}^{\rm EOB}= -0.01670 M\,c^2$ (3PN order).
Note that the fractional difference from 1PN to 2PN order 
is $4\%$ and it increases to $10\%$ from 2PN to 3PN order.
Figure~\ref{fig5} shows that except the very 
recent result of Ref.~\cite{GGB}, which is much closer (probably 
due to their definition of the binary orbital frequency) to the 
PN estimates than the numerical relativity predictions, the 
PN and numerical relativity results~\cite{C,TB,PTC} differ very much. 

\subsection{Beyond the adiabatic approximation: transition inspiral to plunge}
\label{subsec2.3}

Let us now introduce radiation-reaction effects. As discussed in 
Sec.~\ref{subsubsec2.2.1}, Pad\'e approximants~\cite{DIS1} give  
a good estimate of the energy-loss rate ${\cal F}$ 
along circular orbits up to 2.5 PN order. In Ref.~\cite{BD2}
this resummation method was combined with the EOB approach, 
and the authors deduced a system of ordinary differential equations 
which describe the late dynamical evolution 
of a binary--black-hole system. In spherical coordinates 
$(\varphi, R, P_{\varphi},P_R)$, their relevant equations are~\cite{BD2}:
\bea
\frac{d R}{d t} &=& 
\frac{\partial {\cal H}^{\rm impr}_{\rm real}}{\partial P_R}\,, \quad \quad 
\frac{d P_R}{d t} + \frac{\partial {\cal H}^{\rm impr}_{\rm real}}{\partial R}= 0\,, 
\nonumber \\
\frac{d \varphi}{d t} &=& 
\frac{\partial {\cal H}^{\rm impr}_{\rm real}}{\partial P_\varphi}\,, \quad \quad 
\frac{d P_\varphi}{d t} = -\frac{{\cal F}_{\mbox{\small\rm Pad\'e}}(\dot{\varphi})}{\dot{\varphi}}\,.
\label{9}
\eea
Differently from Eqs.~(\ref{2}), the above equations go beyond the 
adiabatic approximation (at least for what concerns the conservative 
part of the dynamics) and can be analytically or numerically  
solved to study the transition between the adiabatic inspiral and the plunge.

Let us discuss briefly the main features of this transition 
in the two extreme limits $\nu \ll 1$ and $\nu = 1/4$ 
[see Ref.~\cite{BD2} for more details]. 
The case $\nu \ll 1$ refers to binary--black-hole systems  
in which a very small black hole spirals around a supermassive 
black hole. They are typical GW sources for the Laser Interferometer Space 
Antenna (LISA). In this case, it was found~\cite{BD1,OT} that the transition 
from adiabatic inspiral to plunge is sharply localized around the ISCO.
Ori and Thorne~\cite{OT} pointed out that likely LISA could observe 
the transition from inspiral to plunge with a signal-to-noise ratio 
of a few. 

For equal-mass binaries ($\nu =1/4$), 
the radiation damping effects are important 
in an extended region on the order of $\Delta (Rc^2/GM) \sim 1 $ 
above the naive (Schwarzschild) ISCO $R=6GM/c^2$. 
The transition from inspiral to plunge is rather blurred 
and the dephasing between the full and the adiabatic 
waveform becomes visible somewhat before the naive ISCO.  
The plunge part of the exact waveform looks like a continuation
of the inspiral part because the orbital motion 
remains quasi-circular throughout the plunge.

\begin{figure}
\begin{center}
{\includegraphics[width=6.5cm,angle=-90]{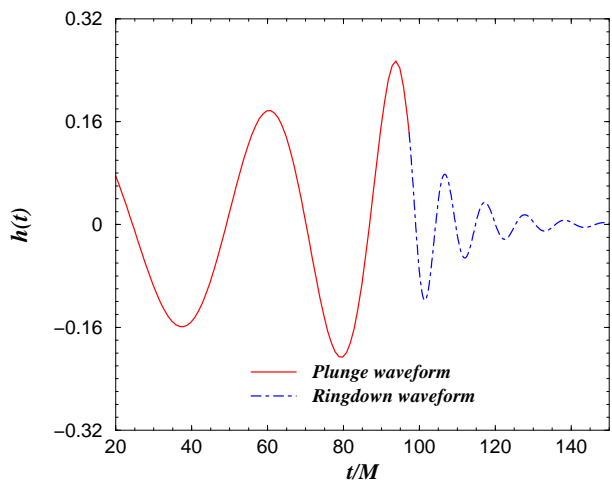}}
\caption{\sl Plunge and ring-down gravitational waveform obtained 
from the EOB approach \protect\cite{BD2} for equal-mass binary.}
\label{Fig7}
\vspace{-0.5cm}
\end{center}
\end{figure}

Figure~\ref{Fig7} shows the plunge, 
merger and ring-down part of the waveform~\cite{BD2} for equal-mass binary. 
The ringdown waveform contains only the mode that is damped more slowly,   
$l=2$, $m=2$~\cite{QNR}, at frequency $\omega_{qnm} \sim 1880 \,
(10 M_\odot/M_{\rm BH})$ Hz, 
where $M_{\rm BH}$ is the mass of the final black hole 
formed. The dimensionless rotation parameter 
is $a_{\rm BH} = J_{\rm BH}/(G M_{\rm BH}^2) = 0.795$, where we denoted 
the angular momentum of the final Kerr black hole by $J_{\rm BH}$.  
The energy emitted during the plunge is $ \sim 0.7 \%$ of $M$, with a comparable
energy loss $ \sim  0.7 \% $ of $ M$ during the ring-down phase~\cite{BD01}. 
This gives a total energy released of $\sim 1.4\%$ of $M$ to be contrasted 
with the much larger value $4-5\%$ of $M$ recently estimated in Ref.~\cite{BBCLT01}.

Recently Damour, Iyer and Sathyaprakash~\cite{DIS3} 
investigated the consequences of the EOB waveform 
for LIGO/VIRGO data analysis. They derived that the 
GW radiation coming from the plunge and merger can significantly enhance 
the signal-to-noise ratio for binaries of total mass 
$M \,\gaq\, 30 M_\odot$. They found that the 
signal-to-noise ratio reaches the maximum value of $\sim 8$ for 
$M \simeq 80 M_\odot$ at $100 \,{\rm Mpc}$. 
Previous estimations using maximally spinning binaries found 
that the merger dominates on inspiral. For example 
Flanagan and Hughes~\cite{FH98} 
predicted that the energy released during the plunge 
is $\sim 10 \% M$ while during the ring-down phase $\sim 3 \% M$ 
and the signal-to-noise ratio reaches the maximum value of $\sim 40$ 
for $M \simeq 200 M_\odot$ at $100 \,{\rm Mpc}$.

\section{Open issues}

\subsection{Spinning binary black holes}
\label{subsec2.4}

The theoretical prediction of GWs from comparable-mass binaries 
is not only affected  by the failure of the PN-expansion, 
but also by spin effects. Various studies~\cite{kidder,apostolatos} 
estimated that if the binary's holes carry a spin, the evolution in time 
of the GW phase will be significantly affected by it -- for example 
the spin can introduce modulations and irregularities 
in the gravitational waveforms. These features can become
quite ``dramatic'' as long as the two spins are big and 
not aligned or antialigned with the orbital angular momentum.  

\begin{figure}
\begin{center}
\epsfig{file=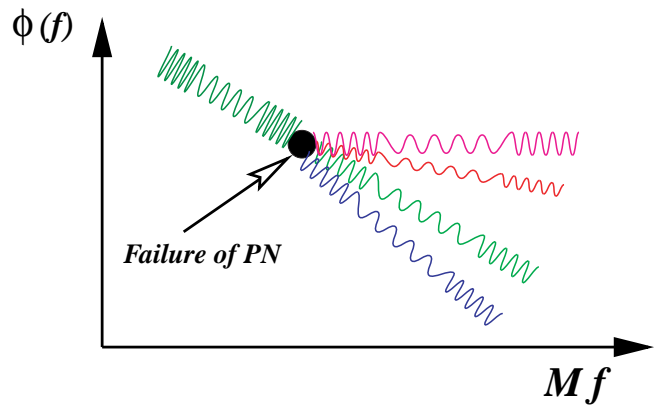,width=0.55\textwidth,angle=0}
\caption{Expected Fourier-domain phase $\phi(f)$ of the GW signal  
versus frequency for various approaches modeling the two-body 
dynamics.}
\vspace{-0.8cm}
\label{fig6}
\end{center}
\end{figure}

To give a rough idea of the effects we are alluding to and that 
we are currently investigating~\cite{BCCV},  
we sketch in Fig.~\ref{fig6} the Fourier-domain phase $\phi(f)$ of the GW signal  
versus frequency for various approaches, 
including possible modulations due to spin effects. 
The various methods give the same prediction for the Fourier-domain 
phase up to the frequency where the PN series fails -- for example for 
$M=20 M_\odot$, $f_{\rm PN-failure} \sim 50 {\rm Hz}$, while 
for $M=30 M_\odot$, $f_{\rm PN-failure} \sim 33 {\rm Hz}$~\cite{BCT98}, 
which are well inside the LIGO/VIRGO band.

More recently, Levin~\cite{levin} claimed that because of 
spin effects the two-body dynamics could be affected 
by chaos, or more in general the dynamics, and as a consequence 
the gravitational waveform, could depend strongly 
on the initial conditions. However, Schnittman and Rasio reanalysed 
this issue in Ref.~\cite{SR}. By calculating the divergence of nearby 
trajectories for a broad sample of initial 
conditions they concluded that the divergent time is much greater 
that the inspiral time. So even if chaos were present it should not 
affect the detection of inspiral waveforms with LIGO/VIRGO.

To tackle these delicate issues it would be desirable to 
extend the resummation methods discussed above to the case 
of spinning binaries. 
To this respect we notice that the Pad\'e resummation method  
has been recently extended to the gravitational flux function 
including spin-orbit and spin-spin effects~\cite{PS}. 
Moreover, on the line of the EOB appoach, Damour~\cite{TD} has 
recently mapped the conservative dynamics of two spinning black holes into the one  
of a test particle moving in an external $\nu$-deformed Kerr metric. 
It will be very important to complete Damour's analysis by including 
radiation reaction effects and describe the late nonadiabatic dynamics of spinning 
black hole binaries.

\subsection{Possible strategy for not missing the GW 
signal from comparable-mass  binary black holes}
\label{subsec2.5}

In sections \ref{subsubsec2.2.1} and \ref{subsubsec2.2.2} we discussed 
the {\em best} theoretical gravitational waveforms currently 
available. These waveforms should certainly be used in detecting and 
extracting physical parameters (masses and spins) for GWs emitted by  
neutron-star binaries and neutron-star/ black hole binaries.  
On the other hand the direct application of these templates  
to detect GWs from comparable-mass black-hole binaries is less straightforward. 
Indeed, the approaches analysed in sections \ref{subsubsec2.2.1} and 
\ref{subsubsec2.2.2} are inevitably affected by approximations and 
assumptions, introduced to overtake the fact that the two-body 
problem is known only up to a certain PN order. 
It would be important for detection purposes to quantify such uncertainties and 
take  them into account when building the best theoretical GW templates. 
If not, the risk is to miss the GW signal. 
A possibility~\cite{BCV} is to {\it deform} and {\it expand} (e.g., 
by introducing new parameters), the template family generated by 
the resummation method having better PN convergent properties, 
in such a way that the new (higher-dimensional) template space could 
(i) include, or at least be not very far, from all the other approaches 
characterized by worse PN convergent properties and (ii) describe 
signals (hopefully the real GW signal!) whose functional form cannot be 
described by any of the original template families. 
To reach this goal the ideas underlying the Fast Chirp Transform 
technique, recently proposed in Ref.~\cite{JP}, deserves 
attention. 
Moreover, for comparable-mass binaries the inclusion of spin effects 
and the enlargement of the template space, hopefully, should not make 
the total number of templates huge, because the number of cycles expected 
in the LIGO/VIRGO band for nonspinning binaries  
is already rather small $\sim 20 -100$. [The situation would be 
different for neutron star or neutron-star/black hole binaries 
for which in the nospinning case we expect a number of cycles 
on the order of $\sim 10^4-10^6$.]

The new template space~\cite{BCV} could be used for on-line search. 
When an acceptable signal-to-noise ratio is found, say $\gaq \, 8$, 
then the output signal could be re-analysed by templates provided, e.g.,  
by the PN Taylor-expanded, Pad\'e approximant and EOB methods, 
to (i) determine which approach better describe the real signal and 
(ii) extract the binary's parameters, as the masses and the spins.

\vspace{-0.5cm}

\ack 

\vspace{-0.5cm}
The author wishes to thank T. Damour, Y. Chen, D. Chernoff, 
B.S. Sathayprakash, K.S. Thorne and M. Vallisneri for 
fruitful collaborations and/or very useful discussions and 
suggestions.

\vspace{0.5cm}
{\it Note added.} While this manuscript was being refereed, an interesting 
paper of Blanchet~\cite{LB} appeared on the web. He 
compares the minimum energy of circular orbits for equal-mass--binary, 
at 3PN order, with the numerical-relativity results of Ref.~\cite{GGB} 
finding good agreement.


\vspace{-0.5cm}

\section*{References}
\begin{thebibliography}{999}
\frenchspacing

\bibitem{Inter} Abramovici A, Althouse W E, Drever R W P, Gursel Y, Kawamura S,
Raab F J, Shoemaker D, Sievers L, Spero R E, Thorne K S, Vogt R E, Weiss R, 
Whitcomb S E
and Zucker M E 1992 {\it Science} {\bf 256} 325;
Caron B et al. 1997 {\it Class. Quantum Grav.} {\bf 14} 1461;
L\"{u}ck H et al. 1997 {\it Class. Quantum Grav.} {\bf 14} 1471;
Ando M et al. 2001 {\it Preprint} astro-ph/0105473.
\bibitem{BCT98} Brady P R, Creighton J D E and Thorne K S 1998 {\it Phys. Rev. D} 
{\bf 58} 061501.
\bibitem{QNR} Chandrasekhar S and Detweiler S 1975 {\it Proc. R. Soc. Lond.} 
{\bf A 344} 441.
\bibitem{postnov} Lipunov V M, Postnov K A and Prokhorov M E 1997 
{\it New Astron.} {\bf 2} 43. 
\bibitem{KT} Thorne K S ``The scientific case for mature LIGO
interferometers,'' (LIGO Document Number P000024-00-R, 
www.ligo.caltech.edu/docs/P/P000024-00.pdf).
\bibitem{PZ99} Portegies Zwart S F and McMillan S L (2000) 
{\it Astrophys. J.} {\bf 528} L17.
\bibitem{DD} Damour T and Deruelle N 1981 {\it Phys. Lett.} {\bf 87A} 81; 
Damour T 1982 C.R. Seances Acad. Sci. Ser. 2 {\bf 294} 1355.
\bibitem{DJS} Jaranowski P and Sch\"afer G 1998 {\it Phys. Rev. D} {\bf 57} 7274;
ibid. 1999 {\it Phys. Rev. D} {\bf 60} 124003;
Damour T, Jaranowski P and Sch\"afer G 2000 {\it Phys. Rev. D} {\bf 62} 044024;
ibid. {\it Phys. Rev. D} {\bf 62} 021501 (R); 
ibid. 2001 {\it Phys. Rev. D} {\bf 63} 044021.
\bibitem{DJSd} Damour T, Jaranowski P and Sch\"afer G 2001 {\it Phys. Lett. B} 
{\bf 513} 147.
\bibitem{DBF} Blanchet L and Faye G 2000 {\it Phys. Lett.} {\bf A271} 58; 
ibid. 2001 {\it J. Math. Phys.} {\bf 42} 4391; 
ibid. 2000 {\it Phys. Rev.  D} {\bf 63} 062005; de~Andrade V C, Blanchet L and Faye 
G 2001 {\it Class. Quant. Grav.} {\bf 18} 753.
\bibitem{1PN} Wagoner R V and Will C M 1976 {\it Astrophys. J.} {\bf 210} 764. 
\bibitem{2PN} Blanchet L, Damour T, Iyer B R, Will C M and Wiseman A G 
1995 {\it Phys. Rev. Lett.} {\bf 74} 3515; Blanchet L, Damour T and 
Iyer B R 1995 {\it Phys. Rev. D} {\bf 51} 536; 
Will C M and Wiseman A G 1996 {\it Phys. Rev. D} {\bf 54} 4813.
\bibitem{2.5PNand3.5PN} Blanchet L 1996 {\it Phys. Rev. D} {\bf 54} 1417; 
Blanchet L 1998 {\it Class. Quantum Grav.} {\bf 15} 113.
\bibitem{BIJ} Blanchet L, Faye G, Iyer B.R., Joguet B 
2002 {\it Phys. Rev. D} {\bf 65} 061501; Blanchet L, Iyer B R and Joguet B 2002 
{\it Phys. Rev. D} {\bf 65} 064005. 
\bibitem{C} Cook G B 1994 {\it Phys. Rev. D} {\bf 50} 5025.
\bibitem{TB} Baumgarte T W 2000 {\it Phys. Rev. D} {\bf 62} 024018.
\bibitem{PTC} Pfeiffer HP, Teukolsky SA and Cook GB 2000 {\it Phys. Rev. D} 
{\bf 62} 104018.
\bibitem{GGB} Gourgoulhon E, Grandcl\'ement P and Bonazzola S (2002) 
{\it Phys. Rev. D} {\bf 65} 044020; Grandcl\'ement, Gourgoulhon E and Bonazzola S 
(2002) {\it Phys. Rev. D} {\bf 65} 044021.
\bibitem{NR} Alcubierre M, Br\"ugmann B, Pollney D, Seidel E and Takahashi R 2001 
{\it Phys. Rev. D} {\bf 64} 061501; Kidder L.E., Scheel M.A. and Teukolsky S.A. 
2001 {\it Phys. Rev. D} {\bf 64} 064017.
\bibitem{BBCLT01} Baker J, Br\"ugmann B, Campanelli M, Lousto C O and 
Takahashi R 2001 {\it Phys. Rev. Lett.} {\bf 87} 121103.
\bibitem{DIS1} Damour T, Iyer B R and Sathyaprakash B S 1998 
{\it Phys. Rev. D} {\bf 57} 885.
\bibitem{BD1} Buonanno A and Damour T 1999 {\it Phys. Rev. D} {\bf 59} 084006. 
\bibitem{BD2} Buonanno A and Damour T 2000 {\it Phys. Rev. D} {\bf 62} 064015.
\bibitem{EOB3PN} 
Damour T, Jaranowski P and Sch\"afer G 2000 {\it Phys. Rev. D} {\bf 62} 084011.
\bibitem{BO} Bender C M and Orszag S A, {\it Advanced Mathematical Methods 
for Scientists and Engineers} (McGraw Hill, Singapore, 1984).
\bibitem{BIZ70} Br\'ezin E, Itzykson C and Zinn-Justin J 1970 {\it Phys. Rev. D} 
{\bf 1} 2349.
\bibitem{KWW} Kidder L E, Will C M and Wiseman A G 1992 {\it Class. Quantum Grav.} 
{\bf 9} L127; ibid. 1993 {\it Phys. Rev. D} {\bf 47} 3281. 
\bibitem{OT} Ori A and Thorne K S (2000) {\it Phys. Rev. D} {\bf 62} 124022.  
\bibitem{DIS3} Damour T, Iyer B R and Sathyaprakash B S 2001 {\it Phys. Rev. D} 
{\bf 63} 044023.
\bibitem{BD01} Buonanno A and Damour T, contributed paper to the IX$^{\rm th}$ Marcel 
Grossmann Meeting (Rome, July 2000); 2000 {\it Preprint} gr-qc/0011052.
\bibitem{FH98} Flanagan E E and Hughes S A 1998 {\it Phys. Rev.  D} {\bf 57} 4535.
\bibitem{kidder} 
Kidder L E, Will C M and Wiseman A G 1993 {\it Phys. Rev. D} {\bf 47} 
4183 (R); Kidder L E 1995 {\it Phys. Rev. D} {\bf 52} 821.
\bibitem{apostolatos} Apostolatos T A, Cutler C, Sussman G J and Thorne K S
1994 {\it Phys. Rev. D} {\bf 15} 6274; Apostolatos T A 1996 {\it Phys. Rev. D} 
{\bf 54} 2438.
\bibitem{levin} Levin J 2000 {\it Phys. Rev. Lett.} {\bf 84} 3515; and 2000
{\it Preprint} gr-qc/0010100.
\bibitem{SR} Schnittman JD and Rasio F 2001 {\it Phys. Rev. Lett.} {\bf 87} 121101.
\bibitem{TD} Damour T 2001 {\it Phys. Rev. D} {\bf 64} 124013. 
\bibitem{PS} Porter E and Sathyaprakash B S, in preparation.
\bibitem{BCCV} Buonanno A, Chen Y, Chernoff D and Vallisneri M, work in progress.
\bibitem{BCV} Buonanno A, Chen Y and Vallisneri M, in preparation.
\bibitem{JP} Jenet F A and Prince T 2000 {\it Phys. Rev. D} {\bf 62} 122001.
\bibitem{LB} Blanchet L 2001 {\it Preprint} gr-qc/0112056.
\end{thebibliography}
\end{document}